\documentclass[prl,twocolumn,superscriptaddress,showpacs,floatfix]{revtex4}

\usepackage{epsfig}
\usepackage{array}
\usepackage{amsmath}
\usepackage{amssymb}
\usepackage{mathbbol}

\newcommand{\taud}{\tau_{\mathrm d}}
\newcommand{\deltaN}{\delta'}
\newcommand{\EF}{E_{\mathrm F}}

\newcommand{\vare}{\varepsilon}
\newcommand{\corrl}{l_V}
\newcommand{\EEE}{(E_1,\widetilde E_1)}
\newcommand{\tm}{\tau_{\mathrm{max}}}
\newcommand{\Et}{(E_1,\tm)}

\begin{document}
\title{Decreasing excitation gap in Andreev billiards\\by disorder scattering}
\author{Florian Libisch}
\email{florian@concord.itp.tuwien.ac.at}
\author{J\"urgen M\"oller}
\affiliation{Institute for Theoretical Physics, Vienna University of Technology, A-1040 Vienna, Austria, EU}
\author{Stefan Rotter}
\affiliation{Department of Applied Physics, Yale University, New Haven, CT
  06520, USA}
\author{Maxim G. Vavilov}
\affiliation{Department of Physics, University of Wisconsin, Madison, WI
  53706, USA}
\author{Joachim Burgd\"orfer}
\affiliation{Institute for Theoretical Physics, Vienna University of
  Technology, A-1040 Vienna, Austria, EU}

\date{\today}


\begin{abstract}

We investigate the distribution of the lowest-lying energy states in
  a disordered Andreev billiard by solving the Bogoliubov-de Gennes equation
  numerically.  Contrary to conventional predictions we
  find a \emph{decrease} rather than an \emph{increase} of the excitation gap
  relative to its clean ballistic limit. We relate this finding to the
  eigenvalue spectrum of the Wigner-Smith time delay matrix between successive
  Andreev reflections. We show that the \emph{longest} rather than the \emph{mean} time
  delay determines the size of the excitation gap. With increasing disorder
  strength the values of the \emph{longest} delay times increase, thereby, in
  turn, reducing the excitation gap.
\end{abstract}

\maketitle

A small metallic grain connected to a superconductor, commonly referred to as
''Andreev billiard'' (AB) \cite{AndreevR}, features very intriguing electron
dynamics that has been the focus of numerous studies, both theoretical and
experimental \cite{stjames,mcmillan, KMG95PRL,BeenPRL67,gueron,altland,GolKup,BelBru,PilBel,
  Elett35MBFB, PRB58LN, MFJ99PRB, PRB82SB, ILVR01, PRB64tarassemchuk, PRL86,
  PRB66CBKV, PRL89AB, CseKor, CrawfordPRE, PRL90SGB, VL,
  PhysRevB.68.220501, KorKau, fytas, libisch:075304, PRB73LRB} (for a review see
\cite{BeenRev}).  The energy spectrum in such a grain is strongly influenced
by the process of Andreev reflection of quasi particles at the contact between
the superconductor and the normal metal. At this ''SN-interface'' an incoming
electron with energy $\vare$ (counted from the Fermi energy $E_F$) is
back-reflected as a ``hole'' with energy $-\vare$ and nearly opposite
momentum~\cite{AndreevR,AbrikosovBook}. Such Andreev reflections result in the
coupling between electron and hole excitations in the normal metal resembling
those of electron-hole states in superconductors. In particular, the density
of states (DOS) near the Fermi edge ($\EF$) is reduced and may exhibit an
``excitation gap'' ($E_1$).  Details of this reduced DOS are determined by the
dynamics in the Andreev billiard which, in turn, depends on the boundary
geometry, the position of the SN interface and on the potential surface in the
AB.  The distance $E_1$ of the first excited state in the grain (``billiard'')
from the Fermi level (set equal to zero in the following) marks the size of
the excitation gap in the energy spectrum. While being much smaller than the
bulk gap $\Delta$ of the superconductor, $E_1\ll \Delta $, $E_1$ may
considerably exceed the mean level spacing $\delta$, i.e.~the average energy
distance between adjacent eigenstates, thus signalling the appearance of a
gap.

\begin{figure}[!tb]
\hbox{}\hfill\epsfig{file=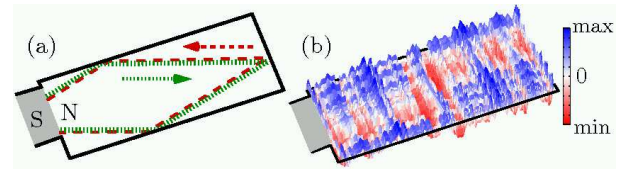, width=8cm}\hfill\hbox{}\caption{(Color online) (a) Geometry of an Andreev billiard (AB) consisting of a
  rectangular normal (N) conductor of dimension ($1.5 W$, $3W$), where $W$ is
  the width of the junction with the superconductor (S). The dotted (dashed)
  line depicts the electron (hole) part of a periodic electron-hole orbit
  created by Andreev reflection at the SN-interface. (b) A sample realization
  of the landscape of the disorder potential inside N.}

\label{fig_rect}
\end{figure}

Qualitative insights into the origin of spectral features in an Andreev
structure, in particular the appearance of a gap, can be gained from a
semiclassical analysis. The semiclassical Bohr-Sommerfeld (BS)
approximation\cite{Elett35MBFB,PRB58LN,PRB66CBKV,ILVR01,libisch:075304} allows
to relate Andreev-reflected periodic orbits with the energy levels of the
Andreev billiard [see Fig.~\ref{fig_rect}(a)]. An energy eigenstate
corresponds to a standing wave along such a periodic electron-hole orbit with
the action difference between electron and hole being quantized. This simple
picture implies that the \emph{lowest} energy $E_1$ in the AB (i.e.~the
excitation gap) will be inversely proportional to the length of the
\emph{longest} electron-hole trajectories.  However instructive the BS
approach may be, it suffers from serious limitations, resulting from the
assumption of strictly retracing electron-hole orbits. Corrections are due to
short-range scattering off disorder~\cite{VL,GolKup,BelBru,PilBel}, quantum
diffraction~\cite{PRB64tarassemchuk,PRL90SGB,PRL89AB,VL,PRB58LN,PRB73LRB},
deviations of the Andreev reflection from perfect
backscattering~\cite{PRL90SGB,PRB73LRB} as well as due to contributions from
trajectories that do not couple to the SN-interface~\cite{PRB66CBKV,PRB73LRB}.
These corrections may turn out to be so large as to render a prediction for
the excitation gap based on the BS approach unreliable. For example, the
formation of a sizeable excitation gap in chaotic Andreev billiards as
predicted by Random Matrix Theory (RMT) and verified numerically
\cite{Elett35MBFB}, is not reproduced by the BS approximation~\cite{PRB82SB}.

In a realistic metal sample brought into contact with a superconductor,
deviations from the ballistic limit by disorder scattering play an important
role.  
If the elastic mean free scattering path $\ell$ is
smaller than the linear dimension of the metal grain, the trajectory between
two successive Andreev reflections at the SN-interface is dominated by
disorder scattering in the interior of the grain rather than by ballistic
scattering off the grain boundaries. It has been suggested that the shortening
of electron-hole orbits or, equivalently, of the average dwell time $\taud$
between successive Andreev reflections by disorder scattering would lead to an
increase of the excitation gap as compared to that of a clean SN
junction\cite{mcmillan,PilBel}. Such a trend would qualitatively be in line
with recent investigations\cite{VL} which have found that the gap in the
ensemble averaged density of states of an AB increases as the mean free path
decreases with respect to the clean, ballistic limit (under the assumption of
constant average dwell times $\taud$).

\begin{figure}
\hbox{}\hfill
\epsfig{file=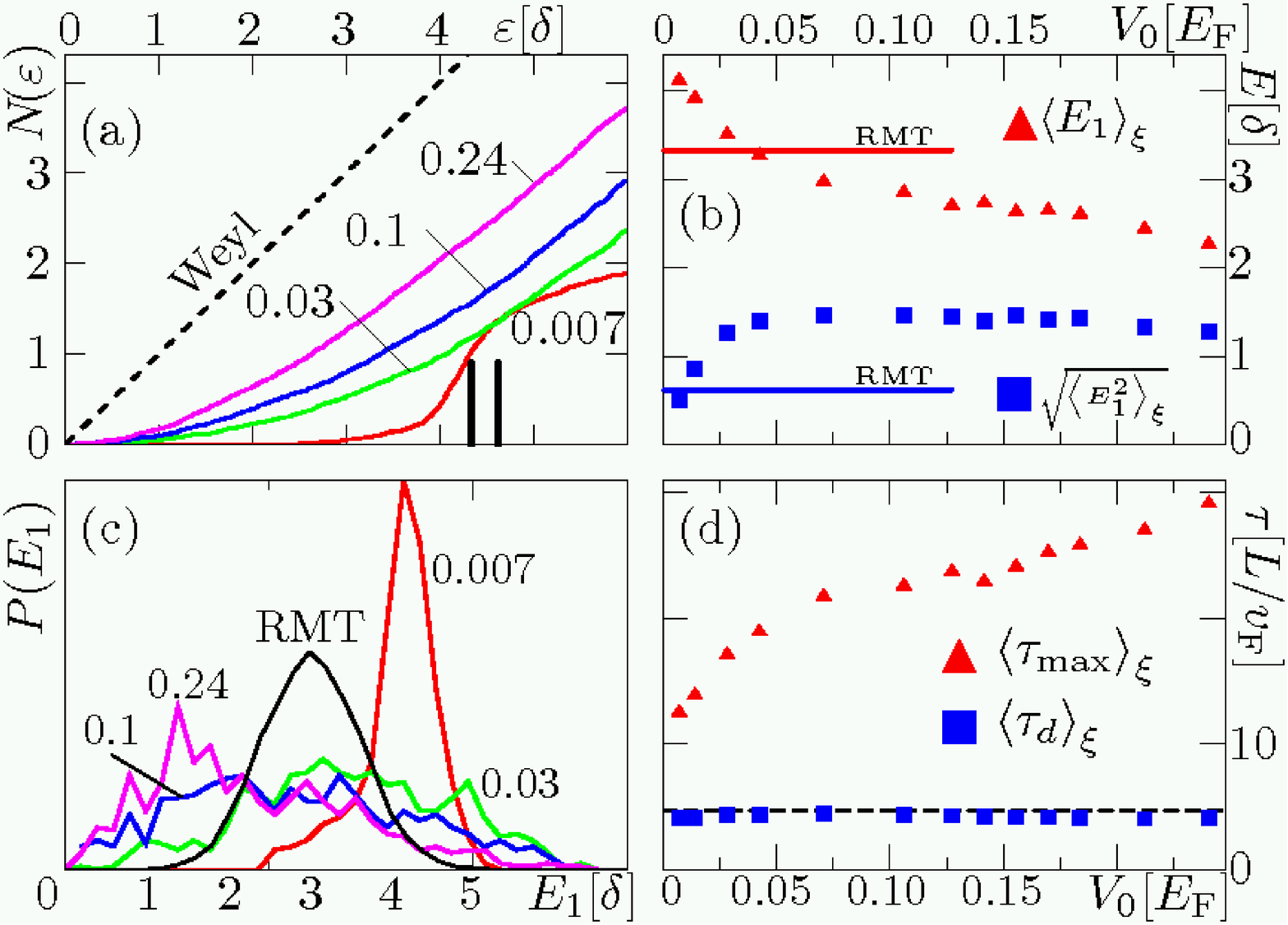,width=8.5cm}
\hfill\hbox{}
\caption{(Color online) (a) Disorder-averaged state counting function,
  $\langle N(\varepsilon)\rangle_\xi$, for four different disorder strengths
  $V_0/E_F=0.007$, $0.03$, $0.1$ and $0.24$ (colored solid lines) and Weyl
  estimate (black dashed line).  The two lowest energy eigenvalues of the
  disorder-free system are marked by vertical bars.  (b) Evolution of the mean
  gap $\langle E_1\rangle_\xi$ (red triangles) and the root mean square
  deviation (blue squares) as a function of disorder strength $V_0$ (in units
  of $\EF$). Horizontal lines mark the RMT predictions. (c) Statistical
  distribution of the lowest eigenvalue $E_1$ for four disorder strengths
  $V_0$ (colored lines) compared with the RMT distribution (black line). (d)
  Dependence of the Wigner-Smith delay times on disorder strength $V_0$. Both
  the mean delay time, i.e., the dwell time $\langle\taud\rangle_\xi$ (blue
  squares), and the maximum delay times $\langle\tm\rangle_\xi$ (red
  triangles) are shown. The black dashed line shows the estimate
    $\langle\tau_d\rangle_\xi=2\pi/N\deltaN$ from \cite{Lyob}.}
\label{fig_eps}
\end{figure}

In this letter, we present numerical \emph{ab initio} simulations for a
two-dimensional AB with disorder, employing the Modular Recursive Green's
function Method (MRGM) in combination with a wave-function matching technique
at the SN interface\cite{PRB62RTWTB,libisch:075304}.
Disorder is represented by elastic scattering off a potential distribution with
short-range disorder with a correlation length $\corrl$ small compared to the
Fermi wavelength, $\corrl/\lambda_F = 0.12$. Decohering processes such as
inelastic scattering are neglected in the following.

We choose a rectangular normal(N)-conducting cavity with
dimensions ($1.5W$, $3W$) where $W$ is the width of the
superconducting lead [see Fig.~\ref{fig_rect}(a)]. We construct
the disorder potential [Fig.~\ref{fig_rect}(b)] by decomposing the
N region into two quadratic modules of dimension ($1.5W$, $1.5 W$)
within each of which we employ a separable random potential,
$V_{{\xi}}(x,y) = V_{\xi_x}(x) + V_{\xi_y}(y)$ [$\xi_x,\xi_y$
denote two different statistical samples, jointly refereed to as
${\xi}\equiv\{\xi_x,\xi_y\}$]. This ``trick'' is employed for
reasons of numerical efficiency, in particular for small
$\lambda_{\mathrm{F}}$ \cite{PRB62RTWTB}. We ensure truly random
scattering by destroying any unwanted separability by rotating by
$180^{\circ}$ the random potentials in the two squares relative to
each other [see Fig.~\ref{fig_rect}(b)].  The spatial correlation
of the random potential is characterized by $\langle
V_{\xi}(x,y)V_{\xi}(x+a,y)\rangle_{x,y}= \langle
V_{\xi}(x,y)V_{\xi}(x,y+a)\rangle_{x,y}= V_0^2\times
\exp(-a/\corrl)$, with $\langle\cdots\rangle_{x,y}$ indicating a
spatial average over the whole disorder area and $\corrl$ the
correlation length.  For a given realization $\xi$ the potential
has zero spatial average, $\langle V_{\xi}(x,y)\rangle_{x,y}=0$,
and an amplitude, $V_0=\sqrt{\langle
  [V_\xi(x,y)]^2\rangle_{x,y}}$\,, which is chosen to be small compared to the
  Fermi energy, $V_0/E_F\lesssim 0.2$.  For $V_0\to 0$ the dynamics in the
  normal conducting part of the AB is entirely ballistic (no disorder
  scattering) and regular (due to the rectangular confinement).
 Calculations are performed with $N=24$
open transverse modes fitting in the lead width $W$ of the superconductor.
The SN-interface itself
is assumed to be ideal (no tunnel barrier) and the superconducting coherence
length is assumed to be smaller than any other length parameters in the
cavity. The superconducting gap $\Delta$ was chosen as $0.2 \EF$ to ensure
that the energy $E_1$ of the lowest-lying eigenstate fulfills $E_1 \ll
\Delta$.

In the fully ballistic limit, i.e., in the absence of any disorder, $V_0=0$,
we find the lowest energy $E_1$ to be four times larger than the AB's mean
level spacing $\delta$ [see Fig.~\ref{fig_eps}(a) for the lowest
eigenenergies]. 
To investigate the influence of the disorder on the energy spectrum we now
gradually increase the disorder amplitude $V_0$. For each value of $V_0$ we
calculate the full energy spectrum (below the superconducting gap $\Delta$)
for 500 different disorder realizations ${\xi}$, and determine the ensemble
averaged state-counting function $N(\varepsilon)$ (i.e., the integrated DOS)
in the ensemble-average.  For very weak disorder strength, $V_0/\EF=0.007$, we
find $\langle N(\varepsilon)\rangle_\xi$ to be still very close to the fully
ballistic limit of $V_0=0$ where the spectral density close to $E_F$ is
strongly suppressed relative to the Weyl estimate $N(\varepsilon) = \rho
\varepsilon$ for the DOS per unit area $\rho=m_{\rm eff}/(\pi\hbar^2)$ [see
Fig.~\ref{fig_eps}(a)].  Increasing the disorder amplitude $V_0$, however,
gradually shifts $N(\varepsilon)$ towards the Weyl distribution [see
Fig.~\ref{fig_eps}(a,b)].  In particular, we find the size of the excitation
gap to be \emph{reduced} with increasing values of $V_0$, rather than
increased.  The reduction of $E_1$ is a sizeable (factor 2 in the range $0\le
V_0/\EF\le 0.2$) and robust effect. For comparison we also show the gap as
predicted by RMT for chaotic systems [see Fig.~\ref{fig_eps}(b)].  These RMT
estimates are based on a numerical calculation representing the internal
dynamics of the normal conductor in the AB by an ensemble of 8000 symmetric
random matrices of size $M\times M$ \cite{PRL86}. $M$ is assumed to scale with
the ratio of cavity circumference $C$ to the size of the SN junction $W$, $M=
N\times C/W$ (for $N=24$ modes in the SN-interface we obtain $M=216$).  While
the RMT value of the gap, $E_1 \approx 3.28\delta$, is in reasonable agreement with
our numerical data for finite disorder strength $V_0 \ne 0$, significant
discrepancies appear for the second moment (i.e.~the variance) of the
distribution $\sqrt{\langle E_1^2\rangle_\xi}$ [see Fig.~\ref{fig_eps}(b,c)].
 The full quantum calculation shows first a steep increase in the
variance with increasing disorder strength before levelling off, whereas the
RMT result underestimates the width of the distribution drastically.  
It should be noted that, strictly speaking, the limit of universality is only
expected to hold for $M\gg N \gg N^{1/3} \gg 1$ \cite{Elett35MBFB}. The latter
limit is difficult to reach in any realistic simulation for a two-dimensional
cavity. The fact
that both the gap size and the variance stay at an almost constant value in a
whole interval of the disorder strength, $0.1\le V_0/E_F\le 0.2$, possibly
points to a saturation effect due to the disorder-induced randomization of
otherwise boundary-specific scattering dynamics. The reduction of gap size and
variance for stronger disorder, $V_0/\EF > 0.2$, may be related to a
transition from weakly disordered scattering to diffusive or localized
dynamics.

The strong \emph{reduction} of the gap size with increasing disorder points to
a mechanism qualitatively different from the behaviour of the mean dwell time
$\langle \taud \rangle_\xi$, which is only negligibly affected by increasing
disorder [see Fig.~\ref{fig_eps}(d)]. To uncover the underlying physics we
employ a rigorous approach that allows to relate the energy spectrum of a
quantum system to the dwell time distribution that does invoke neither any
semiclassical approximation nor random matrix assumptions. Key to our approach
is the relation between the Wigner-Smith (WS) time delay matrix $Q$ and the
scattering matrix S\cite{Wigner55,Smith60},
\begin{equation}
Q(\varepsilon)=-i\hbar S^\dagger(\varepsilon) \partial
S(\varepsilon)/\partial\varepsilon.
\label{eq:Q}
\end{equation}
Equation (\ref{eq:Q}), well-known for unbound quantum systems, can be applied to
 the (bound) spectrum of an AB since an eigenstate of the AB occurs at an energy
$\varepsilon$ for which the determinant\cite{BeenPRL67}
\begin{equation}
\det\!\left[1+S(\varepsilon)S^\dagger(-\varepsilon)\right] = 0,
\end{equation}
where $S(\varepsilon)$ is the scattering matrix of the open, normal-conducting
cavity with the superconductor replaced by a normal conducting waveguide
of identical width $W\!$. Expanding $S(\varepsilon)$ around the Fermi energy
($\varepsilon=0$) for small $\varepsilon$ yields
\begin{equation}
S(\varepsilon)S^\dagger(-\varepsilon) =
\Eins + 2\frac{i}{\hbar}\varepsilon Q + \ldots \approx
\exp\left(2\frac{i}{\hbar}\varepsilon Q\right)\,,
\label{eq:taylor}
\end{equation}
and, in turn, the approximate quantization condition for Andreev
states\cite{CrawfordPRE}:
\begin{equation}
1+\exp\left(2\frac{i}{\hbar}\varepsilon\tau_n\right)= 0\,.
\label{quantization}
\end{equation}
The Wigner-Smith delay times $\tau_n$ are the eigenvalues of $Q$. 
They correspond to ``sticking'' times inside the
normal-conducting cavity between entering and leaving the cavity
through the opening. Since in an AB the opening is replaced by an
SN junction, $\tau_n$ measures the dwell-time between two
successive Andreev reflections. 

\begin{figure}
\hbox{}\hfill\epsfig{file=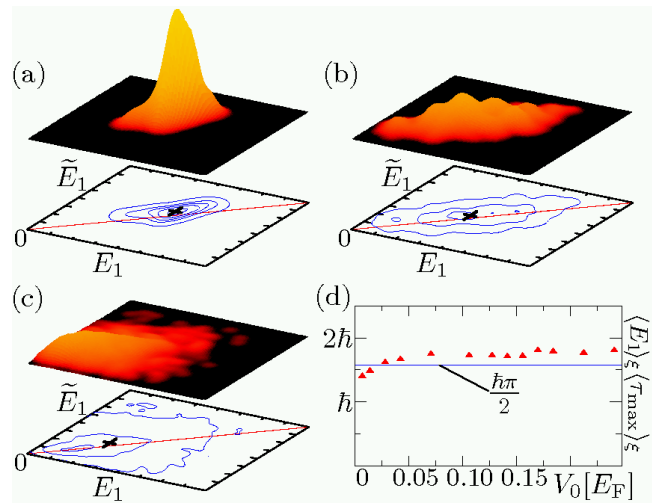, width=8.5cm, clip=true}\hfill\hbox{}
\caption{(Color online) Smoothed distribution of $\EEE$ pairs
  for three different strengths of the disorder potential,(a)
  $V_0/\EF = 0.007$, (b) $0.03$, (c) $0.24$ (500
  realizations of disorder were used). Black crosses in the contour plot mark
  the mean value of the distribution. Perfect $\EEE$ correlation would
  correspond to non-vanishing density only along the diagonal (drawn in the
  contour plot as guide to the eye).  (d) Product of the mean gap
    size $\langle E_1\rangle_\xi$ and the mean of the maximum Wigner-Smith
    delay time $\langle\tm\rangle_\xi$ as a function of disorder
    strength. The constant value $\hbar\pi/2$ [predicted by
    Eq.~(\ref{eq:E1})] is shown for comparison.}
\label{fig_3d}
\end{figure}

The values of 
${\tau}_{n}$ (with $n\le N$) provide a basis-independent measure
for the sticking time of ``quantum trajectories'' without invoking
model-specific assumptions or semiclassical approximations. The only
limitation of Eq.~(\ref{quantization}) is the error of order
$O(\varepsilon^2)$ due to the Taylor expansion 
and approximate resummation of
the unitary operator $S(\varepsilon)S^\dagger(-\varepsilon)$
[Eq.~(\ref{eq:taylor})].
Equation (\ref{quantization}) relates the energy
spectrum at small $\varepsilon$ to
the largest delay-time eigenvalues. In particular, the size of the excitation
gap $E_1$ is determined by the maximal delay time value $\tm$, such that
\begin{equation}
E_1\approx\frac{\hbar\pi}{2\tm}\equiv \widetilde E_1.\label{eq:E1}
\end{equation}
The disorder-averaged maximum delay time, $\langle
\tm\rangle_\xi$, is, indeed, monotonically increasing with
increasing disorder strength $V_0$ [Fig.~\ref{fig_eps}(d)]. In
turn, Eq.~(\ref{eq:E1}) suggests that the disorder-averaged gap
$\langle E_1\rangle_\xi$ will be reduced.

To probe for the correlation between maximum delay time ($\tm$) and the gap
size ($E_1$)
hinted at by Eq.~(\ref{eq:E1}), we have performed a statistical analysis
of the distribution of $\Et$ pairs for 500 disorder realizations
(Fig.~\ref{fig_3d}), converted to $\EEE$ pairs using Eq.~(\ref{eq:E1}). For
perfect correlation we should expect the histogram to feature non-zero bins
only along the diagonal ($\widetilde{E}_1=E_1$). Deviations from a perfect
correlation, resulting in part from the Taylor expansion
Eq.~(\ref{eq:taylor}), provide a measure for the accuracy of the estimate
$\widetilde{E}_1$ as compared to the exact gap 
size $E_1$. For small disorder
strength ($V_0/\EF=0.007$) the correlation between $E_1$ and $\widetilde{E}_1$
is, indeed, near-perfect and non-zero bins occur only in a very limited range
of values $E_1$, $\widetilde{E}_1$ [see Fig.~\ref{fig_3d}(a)]. With increasing
disorder strength [$V_0/\EF=0.03$, see Fig.~\ref{fig_3d}(b)] the maximum in
the distribution shifts to smaller values of $E_1$ and $\widetilde E_1$ while
remaining correlated near the diagonal.  Both observations underscore that
increased disorder decreases the gap which 
is, indeed, correlated with 
the maximum WS time-delay eigenvalue. For much stronger disorder [$V_0/\EF=0.24$, see
Fig.~\ref{fig_3d}(c)], the $\EEE$ correlation is diminished as off-diagonal
bins become more significantly populated. While, on average, the connection
between the disorder-induced reduction of the gap and the increase of the
maximal delay time $\langle\tm\rangle_\xi$ still holds, see
Fig.~\ref{fig_3d}(d), for \emph{individual} strong disorder
realizations this picture breaks down and small (large) gap sizes may well
occur for systems with small (large) values of $\tm$.

\begin{figure}[!b]
\hbox{}\hfill\epsfig{file=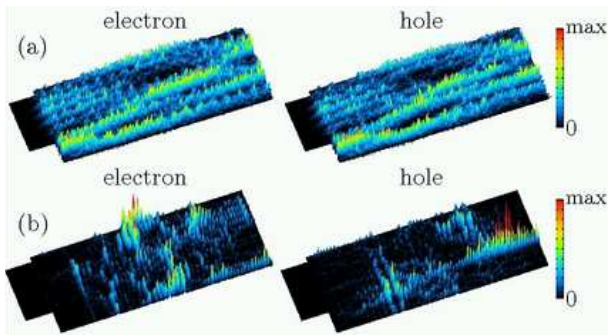, width=8cm, clip=true}\hfill\hbox{}
\caption{(Color online) Electron and hole probability densities of the
  lowest Andreev eigenstate 
  at (a) zero disorder potential, (b) finite disorder strength $V_0/\EF =
  0.15$ (one disorder realization).}
\label{fig_psi}
\end{figure}

The present simulations allow to directly inspect the effect of disorder
scattering on the wavefunction densities in the particle and hole sectors. The
latter provide a microscopic picture of the decay of correlations between gap
and maximum delay time. In the ballistic limit $V_0=0$, the electron and hole
wavefunction tend to closely mirror each other [Fig.~\ref{fig_psi}(a)] in
agreement with retracing electron and hole orbits between two Andreev
reflections. With increasing disorder the similarity between wave components
in the electron and hole sheet gradually disappears [see
Fig.~\ref{fig_psi}(b)]. This observation supports the picture that for strong
disorder the wave function of the lowest AB eigenstate is largely determined
by disorder scattering in the interior rather than by Andreev reflections at
the SN interface. Accordingly, the dwell time between two Andreev reflections
looses significance.

\begin{figure}
\hbox{}\hfill\epsfig{file=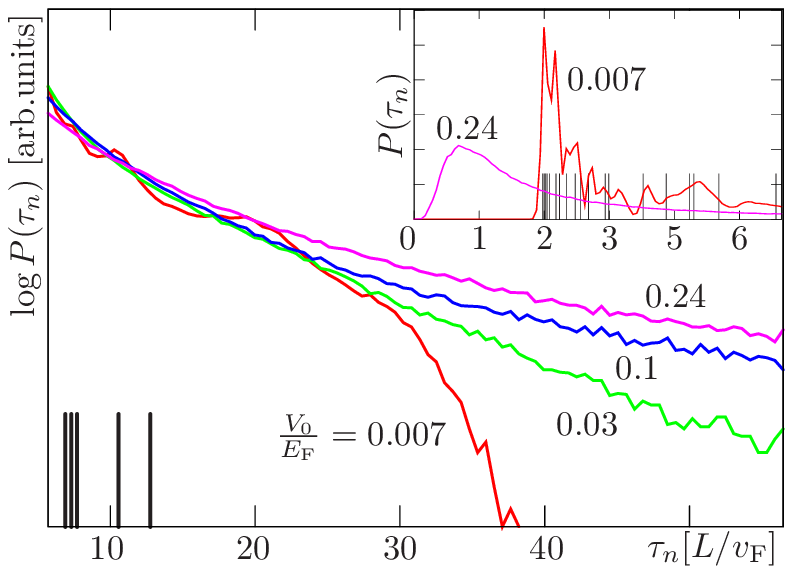, width=8cm}\hfill\hbox{}
\caption{(Color online) Distribution of Wigner-Smith
  delay times, $P(\tau_n)$ (colored lines) for an ensemble average over 500
  disorder realizations $\xi$ and different disorder strength
  ($V_0/E_F=0.007$, $0.03$, $0.1$, $0.24$). For each $\xi$ and $V_0$, the delay
  times were evaluated at 135 different energies in an interval
  $\EF\pm0.1\Delta$. Increasing the disorder strength $V_0$ amplifies the long
  time tail of $P(\tau_n)$ (main plot, logarithmic scale), but concurrently
  produces much shorter delay times (see top right inset).  The disorder-free
  delay times are indicated by vertical bars, with the lowest values starting
  at $\tau\approx 2L/v_F$ (the time of flight across the cavity length $L=3W$ and
  back) and the largest value at $\tau\approx 13L/v_F$. }
\label{fig_tau}
\end{figure}

It is now instructive to inquire into the origin of the discrepancy to those
models suggesting that the presence of disorder induces an increase rather than a
decrease of the gap in comparison to its clean, ballistic limit
\cite{mcmillan,PilBel}.  The key here is the disparate behaviour of the
maximum, $\langle\tm\rangle_\xi\equiv \langle\max_{n=1,N}
\tau_n\rangle_\xi$, and average dwell time,
$\langle\tau_d\rangle_\xi\equiv\langle\sum_n^N\tau_n/N\rangle_\xi$, the latter
of which enters \cite{mcmillan,PilBel}. 
For the system under study here, disorder scattering is obviously
  able to ''delay'' for long-lived scattering states the interval between two
  successive Andreev reflections. The presence of disorder not only
\emph{increases} $\langle\tm\rangle_\xi$, but also \emph{reduces} the minimal
delay time $\langle\tau_{\rm min}\rangle_\xi$ (see Fig.~\ref{fig_tau}).  As a
consequence, the distribution of delay times $P(\tau_n)$ becomes
''stretched'', while leaving the mean value $\langle\tau_d\rangle_\xi$ almost
unchanged. The fact that the average dwell time
  $\langle\tau_d\rangle_\xi$ stays almost unaffected by the increasing
  disorder [see Fig.~\ref{fig_eps}(d)] is in agreement with a general 
  relation\cite{Lyob} between the averaged
  trace of the matrix Q [see Eq.~(\ref{eq:Q})], $\langle \mathrm{tr}\,
  Q\rangle_\xi$, and the mean spacing $\deltaN$ of resonant levels in a
  (normal-conducting) scattering system, $\langle \mathrm{tr}\, Q
  \rangle_\xi=2\pi/\deltaN$ or, equivalently,
  $\langle\tau_d\rangle_\xi=\pi/N\delta$ (note that $\deltaN=2\times\delta$,
  with $\delta$ being the mean level spacing in the corresponding
  AB). Consequently, the mean dwell time should be entirely
  independent of the disorder configuration. This is, indeed, very 
  accurately confirmed by
  our numerical results for $\langle\tau_d\rangle_\xi$ [see
  Fig.~\ref{fig_eps}(d)]. In turn, 
  $\langle\tau_d\rangle_\xi$ is unsuitable for characterizing
  correlations between gap size and disorder strength, because of its 
  independence of $V_0$. Therefore, relating the gap size to the mean dwell
time $\langle\tau_d\rangle_\xi$ also fails to account for the gap reduction
observed here~\cite{footnote}. Clearly, the present results do not preclude an
increase of the excitation gap with increasing disorder for particular
boundary shapes, e.g.~for a gapless excitation spectrum in the absence of
disorder. The present analysis suggests, however, that also in such systems
the behaviour of the longest WS time delay eigenvalue will control the
behaviour of the gap.

The results in Fig.~\ref{fig_tau} demonstrate that for a disordered
  cavity the strength of the disorder ($V_0$) does have a crucial influence on
  the distribution of delay times $P(\tau_n)$ (in particular for long times).
  For chaotic cavities it was found both classically 
  \cite{LinDe} and quantum mechanically \cite{Elett35MBFB,
    brouwerprll,PRB62RTWTB} that the long time tail of delay times does \emph{not}
    depend on certain system-specific parameters as, e.g., the Lyapunov exponent. 
  We therefore expect that the statistical distribution 
  of the gap size undergoes a crossover between the present case of a
  disordered AB and the case of a chaotic AB. It would be interesting to study 
  such a crossover numerically, e.g., by tuning
  the disorder correlation length $l_V$ from the diffractive limit of  
  $l_V\ll \lambda_F$ to the ballistic (chaotic) limit of $l_V\gg \lambda_F$.

With the help of Fig.~\ref{fig_tau}, we can furthermore explain the loss of
correlations among pairs $(E_1,\widetilde E_1)$ for strong disorder
[Fig.~\ref{fig_3d}(c)]: The amplification of the maximal proper delay times
$\langle\tau_{\rm max}\rangle_\xi$ by the increasing disorder translates into
an increase of the high frequency components in the elements of the scattering
matrix $S(\varepsilon)$. As, however, the estimate of the gap size,
$\widetilde{E}_1$, relies in part
 on a Taylor expansion of $S(\vare)$ [see
Eq.~(\ref{eq:taylor})] which can only capture weakly energy dependent
(i.e., low-frequency) components, the accuracy of $\widetilde{E}_1$
deteriorates with increasing disorder strength, thereby gradually diminishing
the correlations among pairs $(E_1,\widetilde E_1)$.  
The behaviour of the mean values $\langle E_1\rangle_\xi$ and
  $\langle \widetilde E_1\rangle_\xi$ can be understood by considering the
  distribution of values $d_E = (E_1\! -\! \widetilde E_1)/2$ (corresponding
  to a projection of the distributions of Fig.~\ref{fig_3d}(a-c) on an axis
  perpendicular to the diagonal $E_1\!=\!\widetilde E_1$). As we have
  verified numerically (not shown), the width of this distribution, 
  $\mathrm{var}(d_E) = \sqrt{\langle d_E^2\rangle_\xi}$, 
  increases with increasing $V_0$,
  while its mean value stays almost unaffected 
  by the disorder strength at $d_E\ll \delta$.
  We speculate that the errors due to the Taylor expansion and the
  resummation of $S^\dagger(\varepsilon) S(-\varepsilon)$ [see
  Eq.~(\ref{eq:taylor})] are randomly distributed and thus cancel out in
  an average over many disorder realizations. 
  This would explain why the averaged values
  $\langle E_1\rangle_\xi$ and $\langle\widetilde E_1\rangle_\xi =
  \langle\tm^{-1}\rangle_\xi$ remain correlated  [see Fig.~\ref{fig_3d}(d)]
 while the correlation between
  individual pairs ($E_1,\widetilde E_1$) breaks down.

In summary, we have numerically calculated the energy spectrum of
electron-hole states in a rectangular Andreev billiard with a tunable disorder
potential. In apparent contrast to qualitative models based on
the mean cavity dwell time $\langle\taud\rangle_\xi$, we find a
\emph{decrease} of the gap size when increasing the disorder amplitude. We
show that this decrease is controlled by the disorder dependence of the
\emph{largest} Wigner-Smith delay time $\tm$ between
subsequent Andreev reflections at the SN-interface. The average
  dwell time $\langle\taud\rangle_\xi$, on the other hand, only depends on the mean level
  spacing, and is thus neither correlated to the evolution of the gap size nor to the
  disorder scattering strength.
Stronger disorder, however, 
drastically increases the value of $\tm$ for long-lived
scattering states. For sufficiently strong
disorder the correlation between the gap size and $\tm$ eventually
breaks down for \emph{individual} disorder realizations, as the eigenenergies
of the system are then more strongly influenced by the specific disorder
potential rather than by the Andreev reflection process. 

\acknowledgments
We acknowledge helpful discussions with
 P.~W.~Brouwer (who also drew our attention to Ref.~\cite{Lyob}), 
J.~Feist and V.~A.~Handara.
Support by the Austrian Science Foundation (FWF Austria, Grant No. FWFP17359), the Max Kade
foundation (New York) and the W. M. Keck foundation is gratefully
acknowledged.


\begin{thebibliography}{36}
\bibitem{footnote} We emphasize here that throughout this letter, we have
used the definition of the gap as the position of the
\emph{lowest} energy state in the AB above the Fermi energy (as
in~\cite{PRL86}). This definition should not be confused with the
gap size in the average density of states, introduced for a large
number of energy levels (as in~\cite{PRB82SB,VL}). 
Also our definition of the dwell time as the mean of the Wigner-smith delay times,
$\langle\tau_d\rangle_\xi=\langle\frac{1}{N}\sum_{n=1}^N\tau_n\rangle_\xi$,
should not be confused with other definitions used in the literature.
\end{thebibliography}

\end{document}